\documentclass[structabstract]{aa}

\usepackage{subfigure}
\usepackage{graphicx}
\usepackage{txfonts}
\usepackage{natbib}
\usepackage{lscape}

\bibpunct{(}{)}{;}{a}{}{,}

\begin{document}

\title{The close Be star companion of $\mathrm \beta$ Cephei}
 
\author{H. E. Wheelwright\inst{1} \and  R. D. Oudmaijer\inst{1} \and  R. S. Schnerr\inst{2}}   

\institute{The School of Physics and Astronomy, EC Stoner Building, The University of Leeds, Leeds, LS2 9JT, UK.\\ \email{pyhew@leeds.ac.uk}\and Institute for Solar Physics, Royal Swedish Academy of Sciences, Albanova University Centre, SE-10691 Stockholm, Sweden.}

\date{Received 08 October 2008 / Accepted 16 February 2009}

\abstract {The {prototype} of the $\mathrm \beta$ Cephei class of pulsating stars, $\mathrm \beta$ Cep, rotates relatively slowly, and yet displays episodic $\mathrm H \alpha$ emission. Such behaviour is typical of a rapidly rotating, classical Be star. For some time this posed a contradiction to our understanding of the Be phenomena as rapid rotation is thought to be a prerequisite for the characteristic emission phases of Be stars. Recent work has demonstrated that the $\mathrm H \alpha$ emission is in fact due to a close companion (separation $\mathrm \approx 0.25\arcsec$) of the star. This resolves the apparent enigma if this close companion is indeed a classical Be star, as has been proposed.}  {We aim to test the hypothesis that this close companion is a valid Be star by determining properties such as its spectral type and $v \sin i$.}  {We employed the technique of {spectroastrometry} to investigate the close binary system. Using the {spectroastrometric} signatures observed, we split the composite binary spectra into its constituent spectra in the \textit{B} band ($\mathrm 4200-5000 \AA$) and \textit{R} band ($\mathrm 6200-7000 \AA$).}  {The {spectroastrometrically} split spectra allow us to estimate spectral types of the binary components. We find that the primary of the close binary system has a spectral type of B2III and the secondary a spectral type of B5Ve. From the relationship between mass and spectral type, we determine the masses of the binary components to be {$\mathrm M_{\mathrm {pri}} = 12.6 \pm 3.2 M_{\sun}$} and {$\mathrm M_{\mathrm {sec}} = 4.4 \pm 0.7 M_{\sun}$} respectively. The {spectroastrometric} data allow some constraint on the orbit, and we suggest a moderate revision to the previously determined orbit. We confirm that the primary of the system is a slow rotator {($v \sin i$ = $\mathrm 29^{+43}_{-29}$ $\mathrm{km\,s^{-1}}$)}, while the secondary {rotates significantly faster}, {at a $v \sin i$ = $\mathrm 230 \pm 45$ $\mathrm{km\,s^{-1}}$}.}  {We show that the close companion to the $\mathrm \beta$ Cephei primary is certainly a valid classical Be star. {It has a spectral type of B5Ve and is a relatively fast rotator}. We confirm that the $\mathrm \beta$ Cephei system does not contradict our current understanding of classical Be stars.}

\keywords{binaries:close -- binaries:general -- Stars:emission-line,Be -- Stars:fundamental parameters -- Stars:individual:$\mathrm \beta$ Cephei}

\titlerunning{The close Be star companion of $\mathrm \beta$ Cep}
\authorrunning{H. Wheelwright et al}

\maketitle

\section{Introduction}

The well-studied object \object{$\mathrm \beta$ Cephei} is a massive,
pulsating star with a spectral type of early B \citep{BSC}. The star
is a prototype of a class of early B type giants and {subgiants} that
exhibit rapid radial velocity, photometric and line profile variations
due to pulsations \citep{Sterken1993}. {$\mathrm \beta$ Cep lies at a
distance of {$d \mathrm = 182 \pm 18$} pc} and has a visual magnitude of
3.2 \citep{hip1997}. {\citet{Donatietal2001} estimate that the star
has a bolometric magnitude of {$L_{bol} = \mathrm -5.8 \pm 0.2$} and a
radius of {$R_* = \mathrm 6.5 \pm 1.2$$\mathrm
R_{\sun}$}}. {Evolutionary tracks suggest that the star's mass is
{$M_*\mathrm = 12 \pm 1$ $\mathrm M_{\sun}$} and that its age is
approximately 12Myr \citep{Donatietal2001}}. As such it seems to be a
fairly typical $\mathrm \beta$ Cep star, as might be expected from the
prototype of the class \citep{Stankov2005}.

\smallskip

In fact, $\mathrm {\beta}$ Cep is a tertiary system. A visual companion
lies at a distance of 13.4\arcsec from the primary
\citep{Heintz1978}. The primary also has a closer, spectroscopic
companion at a separation of approximately 0.25\arcsec
\citep{Gezarietal1972}. Hereafter, we refer to this star as the close
companion and the primary of the system as $\mathrm \beta$ Cep. Little
is known about the close companion. Measurements of the brightness
difference between the primary and the close companion range from 5.0
magnitudes to 1.8 magnitudes
\citep{Gezarietal1972,Balegaetal2002}. \citet{Catanzaro2008}
indirectly estimated the $v \sin i$ of the close companion to be
$\mathrm \approx 230$ $\mathrm{km\,s^{-1}}$ and \citet{Pigulski1992}
estimated its mass to be $\mathrm \approx 8M_{\sun}$. However, using
the same orbital parameters but with a slightly reduced period
\citet{Donatietal2001} conclude the secondary has a mass of
{$M_* = \mathrm 6.2 \pm 0.3M_{\sun}$}.

\smallskip

H$\alpha$ emission emanating from $\beta$ Cep was first reported by
\citet{Karpov1932}, and has since been found both absent and present
with a recurring timescale of approximately a decade
\citep{WilsonSeddon1956,Kaper1995,Panko1997,Neineretal2001}. Such
episodic emission is typical of a classical Be star. However, $\mathrm
\beta$ Cep is a slow rotator with a rotational period of 12 days and a
$v \sin i$ of approximately 25 $\mathrm {km\,s^{-1}}$
\citep{Henrichs2000,Teltingetal1997}. Be stars are typically rapid
rotators rotating at up to 80 \% of their break-up velocity
\citep{PorterandRivinius2003}. As this rapid rotation is thought to be
integral to the H$\mathrm \alpha$ emission of Be stars, the H$\mathrm
\alpha$ emission of $\mathrm \beta$ Cep is not consistent with the
classical Be paradigm. \citet{Donatietal2001} proposed that the Be
behaviour of the star is due to a magnetically confined wind leading
to shocks in the equatorial region ($\mathrm \beta$ Cep does possess
an oblique magnetic field).  However, if this were the case, the
$\mathrm H \alpha$ emission would be modulated by the rotation of the
star, and it is not {\citep{RSSchnerr2006}}. Thus the source of the
$\mathrm H \alpha$ emission has posed a conundrum to our current
understanding of Be stars.

\smallskip

This enigma has been resolved by \citet{RSSchnerr2006}, who used the
technique of {spectroastrometry} to show the $\mathrm H \alpha$
emission is in fact due to the close companion to $\beta$ Cep. This
companion may well be a classical Be star and if this is the case the
contradictory nature of the system to current Be paradigms is
negated. \citet{RSSchnerr2006} used spectra in the R band about
H$\mathrm \alpha$. Here we use {spectroastrometry} to probe the blue
region of the companion's spectrum to test the hypothesis that the
close companion is a classical Be star.

\smallskip

{Spectroastrometry} is a technique which utilises the spatial
information present in a longslit spectrum. The information is
contained in the spatial profile of the spectrum, specifically in the
{photocentre} centroid and the spectral profile's width. Changes in the
flux distribution as a function of wavelength are manifest by changes
in the centroid and width of the spectrum. An unresolved binary system
with one star dominating the flux at an emission line is revealed by a
centroidal displacement towards the dominating star over this
line. Conversely, if one star of the system has a strong absorption
line in its spectrum the centroid of the spectral profile will shift
to the other star over this line. Such signatures can be detected with
high precision, of the order of 1 mas or less
\citep{Oudmaijer2008}. Therefore this is a powerful technique with
which to detect and study close binary systems, as shown by
\citet{DB2006}.

\smallskip

The {spectroastrometric} signature of an unresolved binary system
contains information on the distribution of the flux emanating from
the system. Thus {spectroastrometry} can be not only be used to detect
binary systems, it can also deconvolve the observed spectrum into the
individual spectra of its components. Here we use this technique to
disentangle the spectra of the close $\mathrm \beta$ Cep binary
components to investigate the properties of the close
companion.

\smallskip

This paper is structured as follows: in Sect. \ref{observations} the
details of the observations are presented, in Sect. \ref{results} we
present our {spectroastrometric} results, in Sect. \ref{spec_split} we
discuss the spectra splitting methods and in
Sect. \ref{spec_split_res} we present the results of splitting the
spectra. A discussion and interpretation of the results follows
in Sect. \ref{disscussion} and finally in Sect. \ref{conclusion} we
summarise our findings.

\section{Observations and data reduction}
\label{observations}

\subsection{Observations}

The data presented here were obtained using the 4.2m William Herschel
Telescope (WHT) with the Intermediate Dispersion Spectrograph and
Imaging System (ISIS) spectrograph. The data were obtained on the 7th
of October 2006. Spectra of the $\mathrm \beta$ Cep system in the B
($\mathrm 4200$-$\mathrm 5000 \AA$) and R ($\mathrm 6200$-$\mathrm
6900 \AA$) bands were taken simultaneously using the dichoric slide of
ISIS. The slit width was to set to 5\arcsec. The R1200 and B1200
gratings were used and the resolving power of the spectrograph was
approximately 3800 (measured from telluric lines). The Marconi2 and
EEV12 CCDs were used on the red and blue arm respectively, each with a
pixel size of 13.5$\mathrm{\mu m}$. This resulted in angular pixel
scales of 0.20\arcsec in the blue and 0.22\arcsec in the red
region. As the average seeing during our observations was 1.27\arcsec
the spectral profile was well sampled, which is a requirement for
accurate {spectroastrometry} \citep{JBailey1998a}.

\smallskip

The data were gathered as part of wider study of binary systems. A
wide slit ($\mathrm 5''$) was used to ensure all the light from a
given system entered the slit, despite the effect this had on the
spectral resolution. Multiple spectra were taken at the following
position angles (PA) on the sky: $\mathrm 0^{\degr}$, $\mathrm
42^{\degr}$, $\mathrm68^{\degr}$, $\mathrm 90^{\degr}$,
$\mathrm180^{\degr}$ and $\mathrm270^{\degr}$. Data were taken at a PA
of $\mathrm 42^{\degr}$ and $\mathrm 68^{\degr}$ as
\citet{RSSchnerr2006} suggested the binary was orientated at $\mathrm
\approx 42^{\degr}$. Dispersion calibration arcs were made using CuNe
and CuAr lamps.

\subsection{Data reduction}

Data reduction was conducted using the Image Reduction and Analysis
Facility (IRAF)\footnote{IRAF: written and supported by the IRAF
programming group at the National Optical Astronomy Observatories
(NOAO) in Tuscon Arizona (http://iraf.noao.edu/))} and routines
written in Interactive Data Language (IDL). Flat field and bias frames
were combined and the averaged flat field was then normalised. The raw
data were then corrected using the averaged bias frame and the
normalised average flat frame. Saturated exposures were discarded. The
total intensity {longslit} spectra were then extracted from the
corrected data in a standard fashion. Wavelength calibration was
conducted using the arc spectra taken after the science observations
at a position angle of $\mathrm 270^{\degr}$.

\smallskip

{Spectroastrometry} was performed by fitting Gaussian profiles
to the spatial profile of the {longslit} spectra at each
dispersion pixel. Spurious fits (for example due to cosmic rays) were
identified and {discarded, allowing the routine to fit the
spectral profile}. This resulted in a positional spectrum -- the
centroid of the Gaussian as a function of wavelength -- and a FWHM
(Full-Width-at-Half-Maximum) spectrum -- the FWHM as a
function of wavelength. The continuum position exhibited a general
trend across the CCD chip (of the order of 10 pixels). This was
removed by fitting a low order polynomial to the continuum regions of
the spectrum. This set the continuum position of the centroid to
zero. Spot checks were used to ensure line effects were not fit by the
function.

\smallskip

 A correction for slight changes in the dispersion was determined by
{cross correlating} individual intensity spectra, and was then applied
to the associated intensity, positional and FWHM spectra. This was to
ensure slight changes in wavelength (due to flexure of the
spectrograph) did not introduce spurious signatures when spectra
obtained at differing position angles were combined. All intensity,
positional and FWHM spectra at a given position angle were then
combined to make an average spectrum for each position angle.

\smallskip

 The average positional spectra for anti-parallel position angles were
 combined to form the average North-South (NS) and East-West (EW)
 positional spectra, i.e.: ($\mathrm 0^{\degr} - 180^{\degr}$)/2 and
 ($\mathrm 90^{\degr} - 270^{\degr}$)/2. This procedure eliminates
 instrumental artifacts as real signatures rotate by $\mathrm
 180^{\degr}$ when viewed at the anti-parallel position angle while
 artifacts remain at a constant orientation. However, in some of the
 data a signature was noted at only one position angle. Subtraction of
 the two anti-parallel spectra would not remove this effect but
 clearly it is also an artifact. Thus all average position spectra
 were assessed visually to exclude features only present at a single
 position angle. Signatures in data taken at position angles with no
 anti-parallel counterpart were judged real if they exhibited
 qualitative similarity to signatures at a similar position angle that
 were judged real. The FWHM spectra at opposite position angles were
 visually inspected to search for artifacts. Features present in data
 taken at only one position angle were classified as an artifact and
 discarded. {FWHM} spectra at anti-parallel position angles were combined to  make an average NS or EW FWHM spectrum.

\section{Results: The {spectroastrometric} signatures observed}
\label{results}

\subsection{The spectrum of $\mathrm \beta$ Cep}

\begin{center}
\begin{figure*}
\centering
\begin{tabular}{c}
{\includegraphics[scale=0.75]{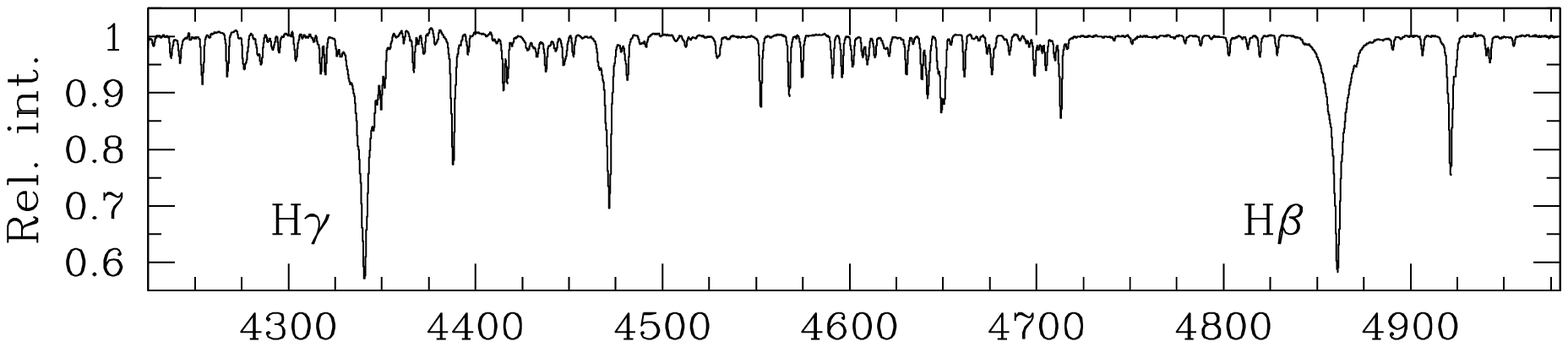}} \\
{\includegraphics[scale=0.75]{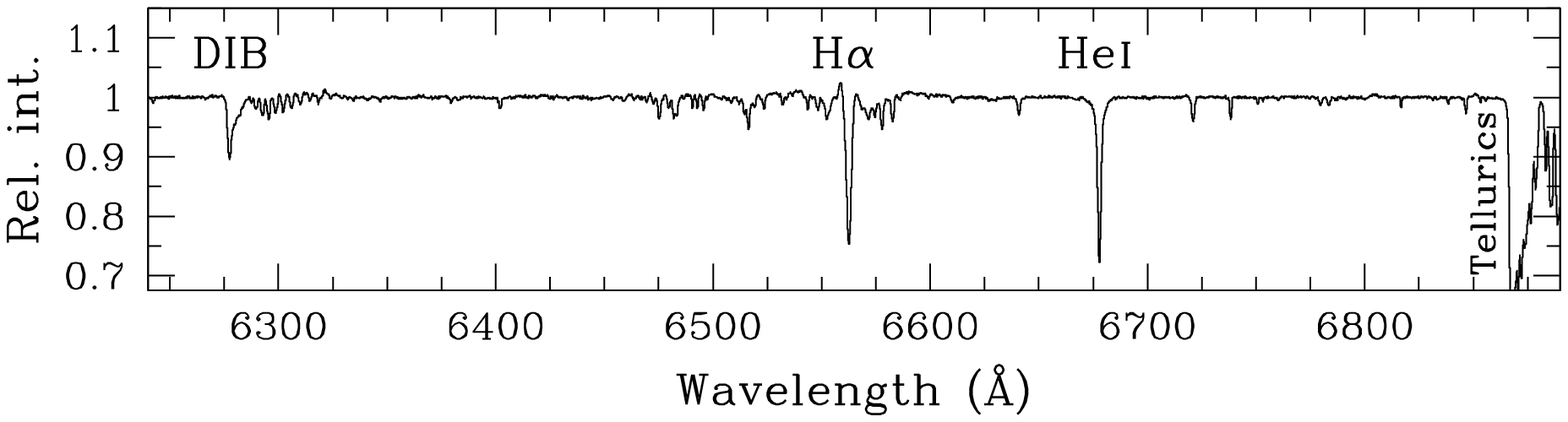}} \\
\end{tabular}
\begin{center}
\caption{The average spectra of $\mathrm \beta$ Cep (total light of the binary system) in the blue (upper panel) and the red (up to the telluric lines -- lower panel) spectral range.}
\label{total_spec}
\end{center}
\end{figure*}
\end{center}

In Fig.\ \ref{total_spec} we present the average spectra observed in
both the \textit{B} and \textit{R} range. H$\mathrm \alpha$ can be
seen in absorption with a small double peaked emission profile, with
the blue peak just rising above the continuum level. The hydrogen
Balmer line profiles are also presented in the top panels of
Fig.\ \ref{spec_ast_plots}. No emission is noticeable in the H$\mathrm
\beta$ and H$\mathrm \gamma$ lines. Besides $\mathrm H\alpha$ other
prominent lines in the red region are \ion{He}{i} at $\mathrm 6678
\AA$ and a Diffuse Interstellar Band (DIB) at $\mathrm \approx 6280
\AA$. In the blue region many more absorption lines are present, most
prominent of which are the lines due to \ion{H}{i} and \ion{He}{i}. The
numerous weaker and narrow lines present are primarily due to
\ion{O}{ii}, \ion{Si}{ii}, \ion{Si}{iii} and other ionised
metals such as \ion{C}{ii} and \ion{N}{ii}.

\subsection{The {spectroastrometric} results}

\begin{center}
\begin{figure*}
\centering
\begin{tabular}{c c c}
  {\includegraphics[scale=0.3]{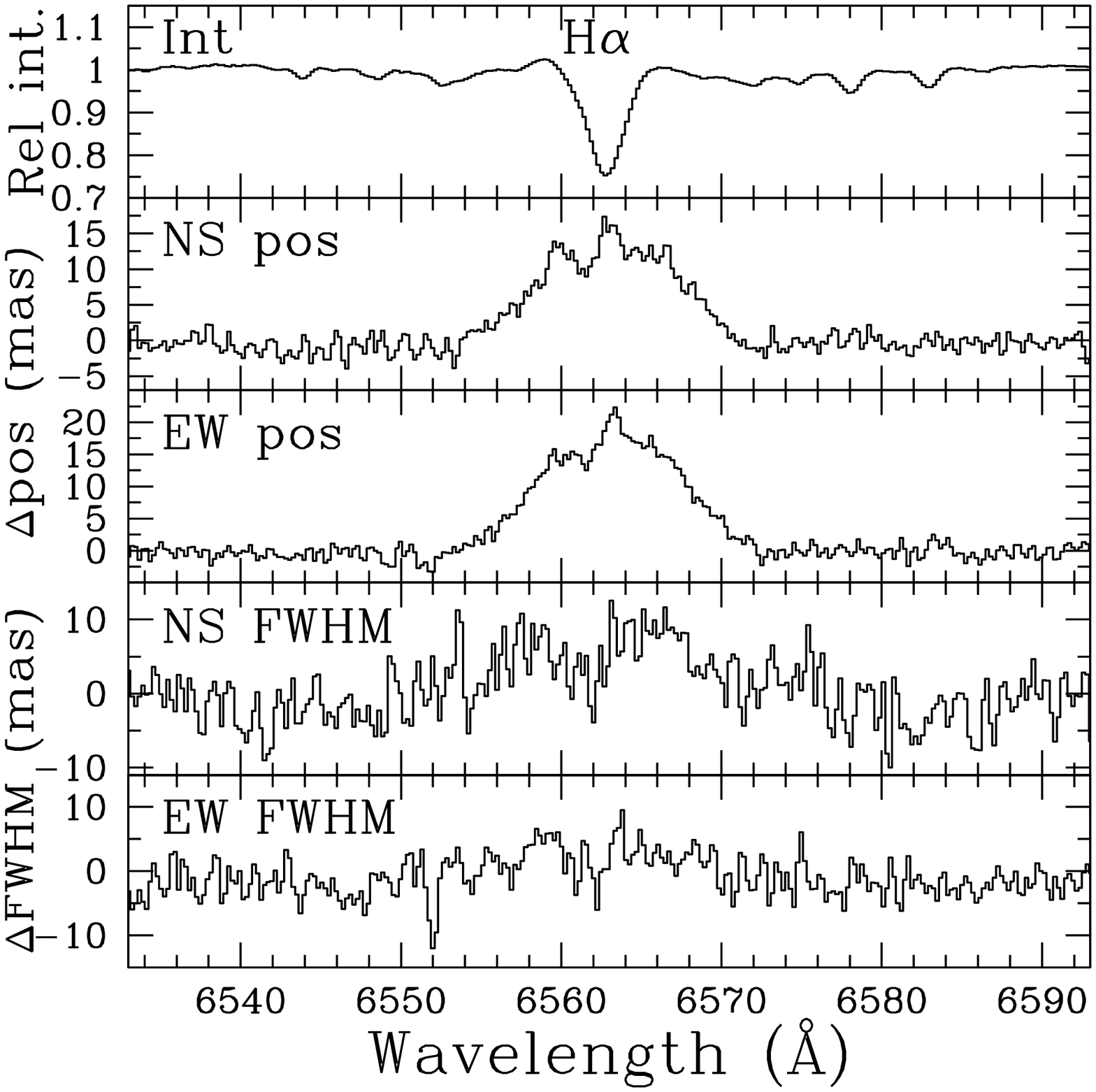}} & 
  {\includegraphics[scale=0.3]{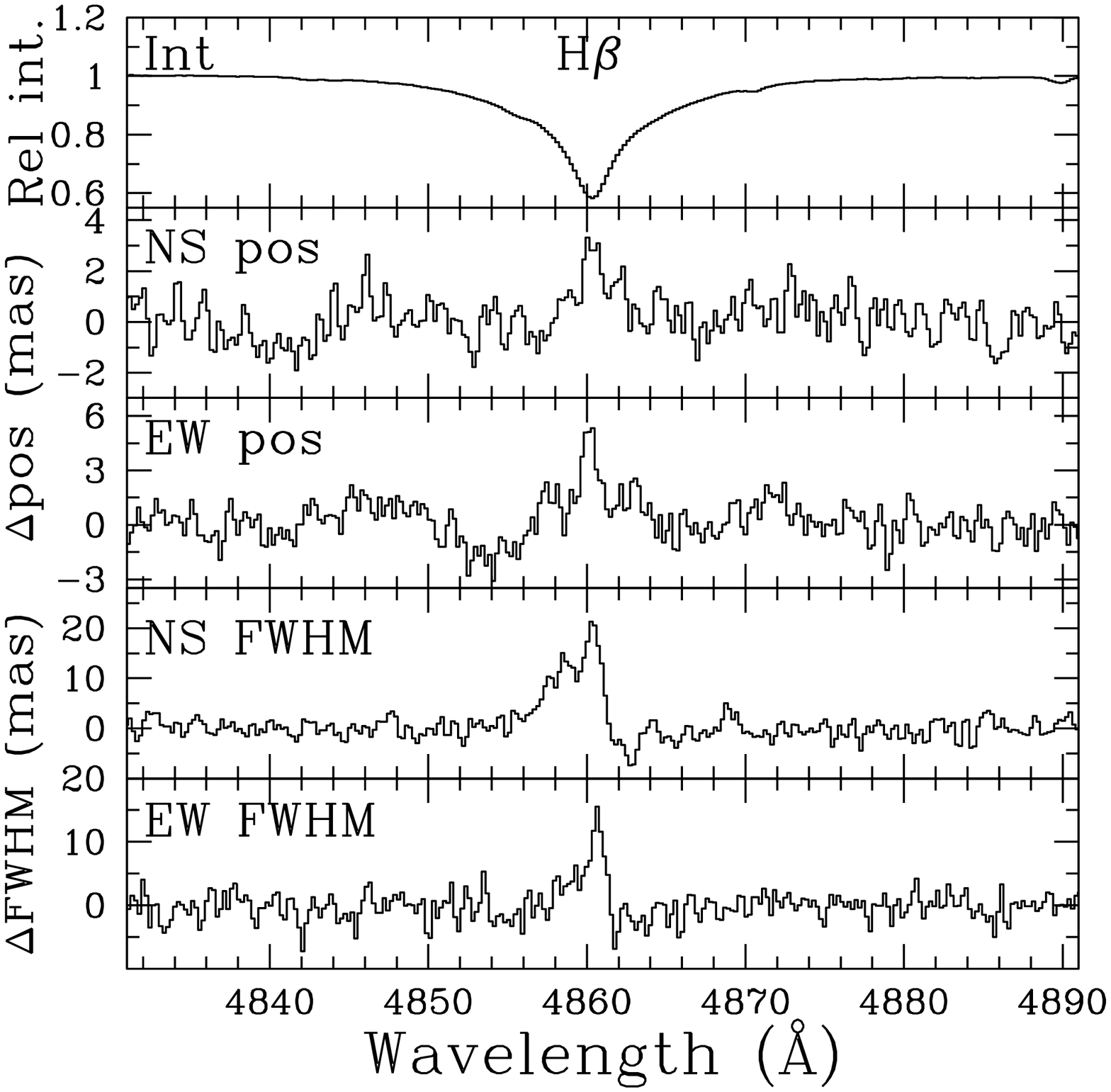}} &
  {\includegraphics[scale=0.3]{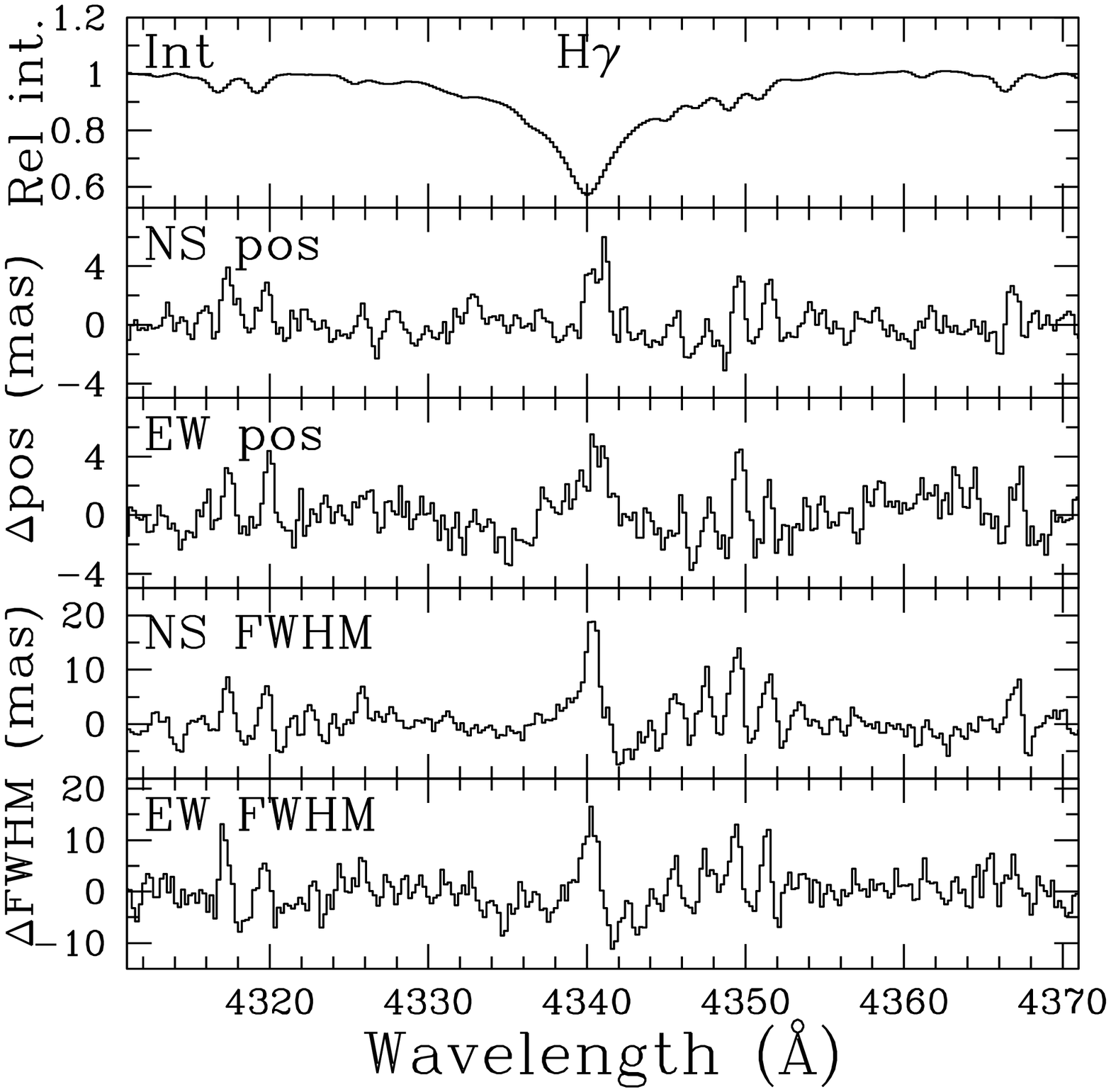}} \\
\end{tabular}
\begin{center}
\caption{The {spectroastrometric} signature of the close $\mathrm \beta$ Cep binary system across the $\mathrm H \alpha$, H$\mathrm \beta$ and H$\mathrm \gamma$ lines (from left to right). Across each line
we present the averaged intensity profile normalised to the continuum,
the position of the {photocentre} with respect to the continuum
position in the NS and EW direction and the FWHM of
the spectral spatial profile, with respect to the continuum value, in
the same directions. In the positional spectra North and East are
presented as being in the positive direction (i.e. to the top of the
page).}
\label{spec_ast_plots}
\end{center}
\end{figure*}
\end{center}

The {spectroastrometric} signatures of the close $\mathrm \beta$ Cep
 binary system over the three principle hydrogen Balmer lines are
 presented in Fig.\ \ref{spec_ast_plots}. For each line we present:
 the intensity profile and the centroidal signature and change in the
 FWHM of the longslit spectra over these lines in the NS and EW
 directions. The {photocentre} of the spectral profile shifts to the
 North and East and the FWHM increases over these lines. This means we
 clearly detect the close binary system {--} something not trivial in
 seeing conditions of greater than 1\arcsec. The signature is most
 prominent over the $\mathrm H \alpha$ line while small and consistent
 features can be noted across H$\mathrm \beta$ and $\mathrm H
 \gamma$. The {spectroastrometric} signature observed over the emission
 component of the H$\mathrm \alpha$ profile implies that the source of
 the emission lies to the North-East (NE) and the emission profile is
 intrinsically broader than the absorption profile. This confirms the
 result of \citet{RSSchnerr2006}, i.e. the H$\mathrm \alpha$ emission
 is emanating from the companion in the NE and not the primary. 

\smallskip

Across the $\mathrm H \beta$ and H$\mathrm \gamma$ lines the
 positional excursions occur in the same direction as over H$\mathrm
 \alpha$. As these signatures take place across an absorption line
 this indicates that the South-West (SW) component of the system
 'dominates' the absorption profile. In addition narrow positional
 excursions to the NE and FWHM increases can be seen over the lines
 seen alongside H$\mathrm \gamma$ (primarily \ion{O}{ii}
 lines). This indicates that these lines are associated with the SW
 component of the system, the primary. These data highlight the
 exquisite sensitivity of {spectroastrometry} to changes in the flux
 distribution. Changes in the {photocentre} of the order of 5 mas or  less are detected. The noise in the positional spectra is typically
 of the order of 1 mas.

\smallskip

\begin{center}
\begin{figure*}
\centering
\begin{tabular}{c c c}
{\includegraphics[scale=0.25]{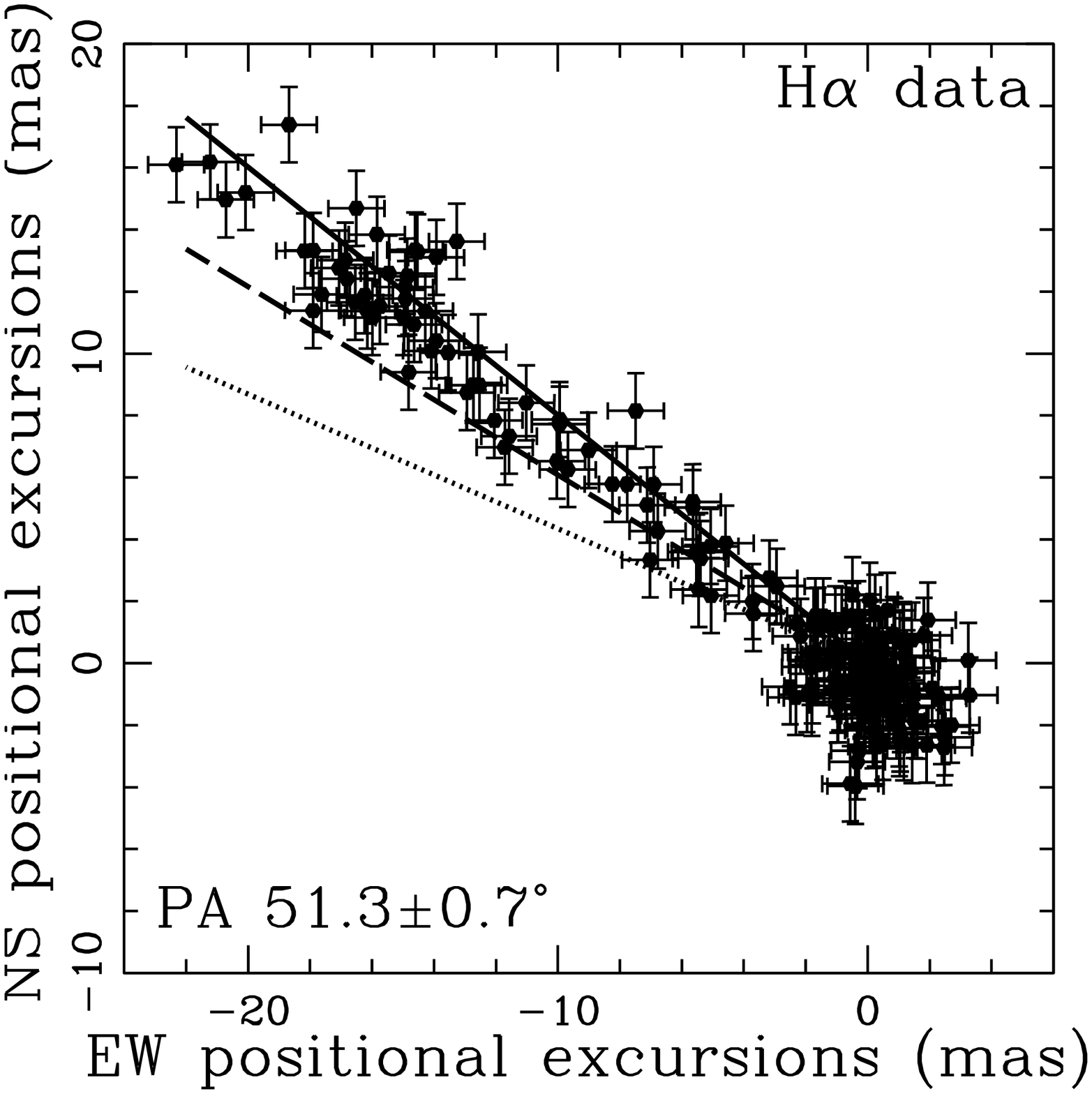}} &
{\includegraphics[scale=0.25]{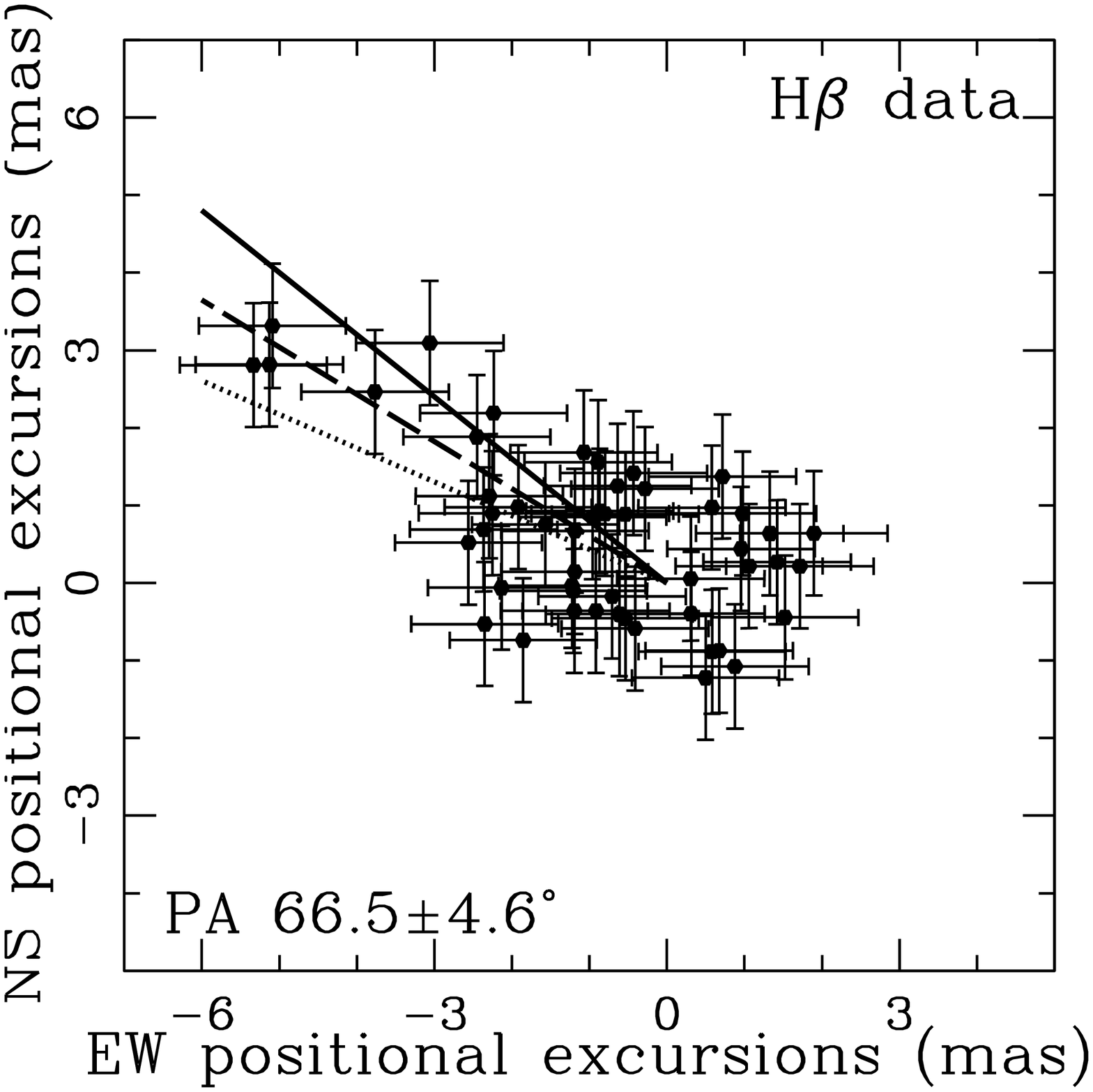}} &
{\includegraphics[scale=0.25]{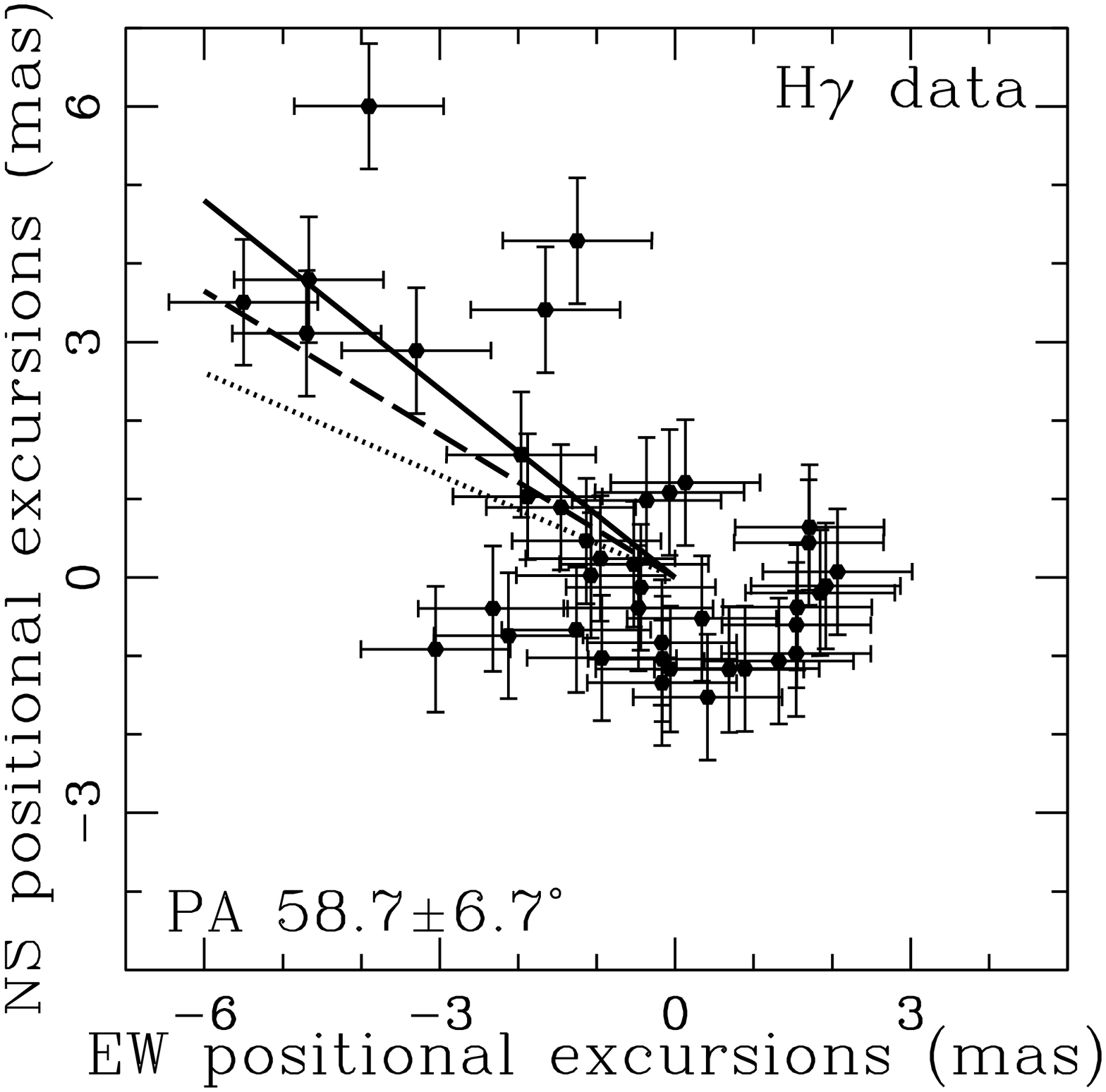}} \\ 
\end{tabular}
\begin{center}
\caption{XY plots of the {spectroastrometric} signature of the close $\mathrm \beta$ Cep binary system over the H$\mathrm \alpha$, H$\mathrm \beta$ and H$\mathrm \gamma$ lines. The solid line is a fit to the H$\mathrm \alpha$ data, the dotted line is a fit to the H$\mathrm \beta$ data and the dashed line a fit to the H$\mathrm \gamma$ data. North is to the top of the page and East is to the left.}
\label{xyplot}
\end{center}
\end{figure*}
\end{center}

The 'XY plots' (NS against EW excursions) of the $\mathrm \beta$ Cep
system are presented in Fig.\ \ref{xyplot}. The XY plots trace a
straight line to the North and East {--} the direction in which the
source of the H$\mathrm \alpha$ emission lies.  The position angle of
the binary is determined from a simple least-squares fit to the data, and is
in general consistent across the different lines ($\mathrm
H\alpha$: $\mathrm51.3 \pm 0.7\degr$, $\mathrm H\beta$: $\mathrm 66.5
\pm 4.6\degr$ \& $\mathrm H\gamma$: $\mathrm 58.7 \pm
6.7\degr$). However, the H$\mathrm \beta$ position angle is not
consistent with the H$\mathrm \alpha$ data. The H$\mathrm \beta$ and
H$\mathrm \gamma$ lines were associated with much weaker positional
excursions and as such these position angles are more uncertain than
the position angle derived from the H$\mathrm \alpha$ data.

\section{Splitting the spectra}

\label{spec_split}

Two approaches were used to split entangled/composite binary
spectra. The first approach was pioneered by
\citet{JBailey1998a}. This method utilises the fact that the changes
in the flux ratio of a binary system lead to positional displacements
of the {photocentre}. The movement of the {photocentre}
across a given wavelength is proportional to the component separation
and the component flux ratio at the particular wavelength. This
centroid movement takes place about the continuum position, which is
determined by the continuum brightness ratio of the two
components. Therefore if the separation and the continuum flux ratio
of the two components are known the intensity spectra and positional
spectra observed can used to disentangle the individual fluxes of the
two components.

\smallskip

The second approach, that of \citet{JMPorter2004}, does not require
any prior knowledge of the binary system. This method not only
deconvolves spectra, it also estimates the separation of the two
binary components. This method is based upon numerical simulations of
point source binary systems performed by
\citet{JMPorter2004}. \citet{JMPorter2004} determined the dependence
of the {spectroastrometric} signature of a given binary on the system's
properties. Using the relationships established by these simulations
one can use the three {spectroastrometric} observables (the centroid,
total flux and width at a given $\mathrm \lambda$), with knowledge of
the seeing, to recover the individual fluxes of the binary
components. For a full description of the method see
\citet{JMPorter2004}.

\smallskip

We have used both methods for two reasons. Firstly there are certain
limitations to the method of \citet{JMPorter2004}. We found accurate
knowledge of the seeing was essential, something not trivial in this
case. To test the reliability of the method of \citet{JMPorter2004} we
applied this method on simulated data. It was found that with accurate
knowledge of the seeing this method consistently returned the
separation of the two point sources and their respective fluxes, to
within an error of $\mathrm \approx 10 \%$ or better. However, when
adopting the minimum FWHM as an estimate of the seeing (which is
actually an upper limit) the method can under estimate the
separation. While we expect this effect to be small we treat our
estimate of $\mathrm \sigma_{s}$ with caution and use the method of
\citet{JBailey1998a} as a consistency check.

\smallskip

Secondly, we do not just use the method of \citet{JBailey1998a} as
although we can find literature values for the difference in
brightness of the two components and their separation, none are
certain. The difference in magnitude reported ranges from 1.8 to 5.0
\citep{Balegaetal2002,Gezarietal1972} while the separation changes
slowly due to the motion of the binary system. The orbit has been
determined by \citet{Pigulski1992}. However this orbit is not
consistent with our results and those of \citet{RSSchnerr2006} (see
Sect. \ref{beta_cep_orbit}). Therefore, we do not use the orbital
parameters of \citet{Pigulski1992} to estimate the position of the
secondary. Thus we used each method as a consistency check upon the
other.

\subsection{The method of Bailey (1998)}  
\label{bailey_spec_split}

The method of \citet{JBailey1998a} uses two
{spectroastrometric} observables (centroidal excursions and the
total flux observed) and two inputs (binary projected separation and
difference in brightness). Three values for the difference in
brightness between the binary components were taken from the
literature: 1.82 magnitudes at $\mathrm \lambda \approx 8100 \AA$, a
rough average value of 3.4 magnitudes around $\mathrm 5500 \AA$ and 5
magnitudes at $\mathrm 5000 \AA$
\citep{Balegaetal2002,Hartkopfetal2001,Gezarietal1972}. Distances in
the literature between the primary and the secondary range from
$\mathrm \approx 0.25\arcsec$ to 0.04\arcsec.  The most recent
observation of the system was in 1998, and thus the position of the
secondary in 2006 is uncertain.

\smallskip

We used a range of distances (0.05\arcsec to 0.25\arcsec) with each of
the magnitude differences listed above, and evaluated the results
based on the following criteria. Any input parameters
leading to negative flux in the secondary spectra were immediately
disregarded.  Also situations where absorption lines in the primary
were mirrored by an emission line in the secondary spectra were
considered unlikely.  

\smallskip

In the red region a brightness difference of 5 magnitudes was
discarded as it resulted in secondary spectra with negative flux. A
difference in brightness of 3.4 magnitudes was discounted as this led
to secondary spectra with \ion{He}{i} $6678 \AA$ in emission, an
uncommon occurrence in field stars. Using $\Delta R$ of 1.8 with a
range of separations did not result in a significant constraint upon
the separation of the two components. A separation of 0.05\arcsec was
discarded as this resulted in \ion{He}{i} emission in the secondary
spectrum. Separations of 0.1\arcsec  to 0.15\arcsec  resulted in a
secondary spectrum devoid of any \ion{He}{i} $\mathrm 6678 \AA$
feature while separations of 0.20\arcsec  and 0.25\arcsec  led to a
secondary spectra with a \ion{He}{i} $\mathrm 6678 \AA$ absorption
feature. If the secondary is a mid B type star the presence of
\ion{He}{i} 6678$\rm\AA$ is to be expected, but the absence of a
\ion{He}{i} feature is consistent with a late type B star. Both
spectral types are possible and thus the binary separation is not
constrained beyond the range 0.05\arcsec  and 0.25\arcsec.

\smallskip

\begin{center}
\begin{figure*}
\begin{center}

{\includegraphics[scale=1]{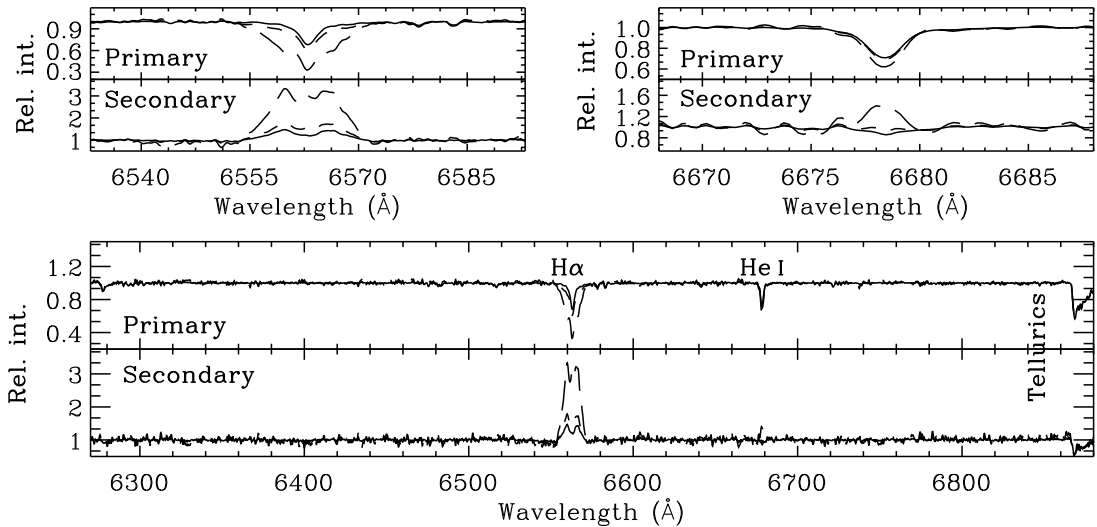}}\\

\end{center}
\begin{center}
\caption{The red spectrum of the close $\mathrm \beta$ Cep binary system, split using the method \citet{JBailey1998a}, $\mathrm\Delta R$ of 1.8 and a range of binary separations: 0.05\arcsec (long-dashed line), 0.15\arcsec (short-dashed line) and 0.25\arcsec (solid line). In the top left the wavelength range is reduced and centred on $\mathrm H \alpha$ while in the top right attention is focused upon the \ion{He}{i} $\mathrm 6678.15 \AA$ profile.}
\label{red_spec_bailey}
\end{center}
\end{figure*}
\end{center}

In Fig.\ \ref{red_spec_bailey} we present the average red spectra of
the close $\mathrm \beta$ Cep binary system, split using the method of
\citet{JBailey1998a}, a brightness difference of 1.8 magnitudes and
separations of $\mathrm 0.05''$, $\mathrm 0.15''$ and $\mathrm
0.25''$.  It is evident that the H$\mathrm \alpha$ emission is
associated with the secondary. The {double peaked} H$\mathrm \alpha$
emission profile is typical of rapidly rotating classical Be stars. 

\begin{center}
\begin{figure*}

\begin{center}
\begin{tabular}{c}
{\includegraphics[scale=1]{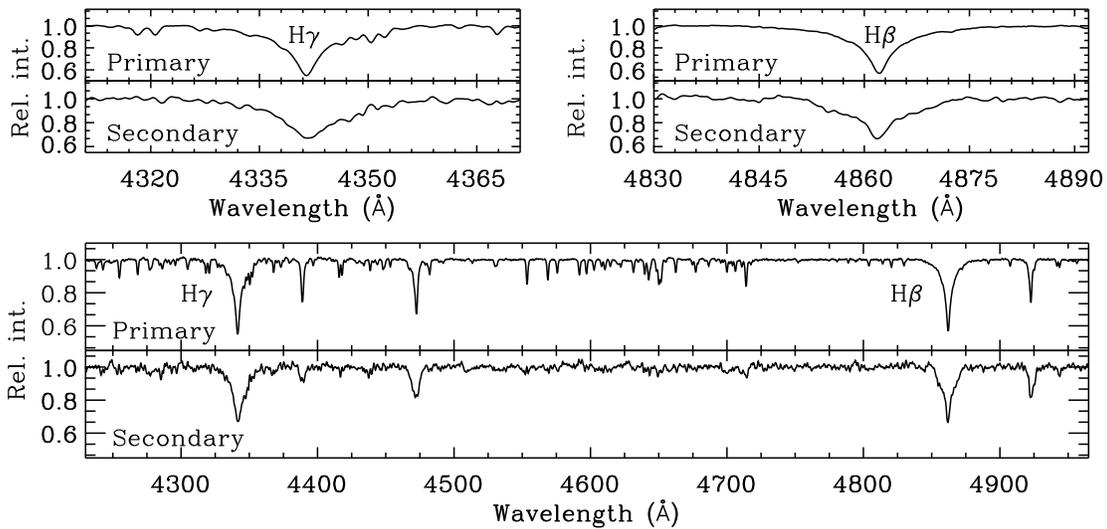}}\\
\end{tabular}
\end{center}
\begin{center}
\caption{The blue spectrum of the close $\mathrm \beta$ Cep binary system, split using the method \citet{JBailey1998a}, $\mathrm \Delta$ B = 1.8 and d = 0.20$\mathrm''$. In the top left the wavelength range is reduced and centred on $\mathrm H \gamma$ while in the top right attention is focused upon the H$\mathrm \beta$ profile.}
\label{blue_spec_bailey}
\end{center}
\end{figure*}
\end{center}

When splitting the blue spectra differences in brightness between the
two binary components of 5.0 and 3.4 were discarded due to similar
arguments as presented above. Thus it appears the brightness
difference between the two binary components in the \textit{B} band is
approximately 1.8 magnitudes, as in the \textit{R} band. In Fig.\
\ref{blue_spec_bailey} we present the spectra obtained with this value
of $\mathrm \Delta$B and a separation of 0.20\arcsec. {The many
narrow absorption lines present in the composite spectrum
(\ion{O}{ii}, \ion{Si}{iii} etc.) are clearly associated with the
primary spectrum. Also evident is that the \ion{He}{i} absorption
lines are weaker in the secondary spectrum than in the primary
spectrum.} The secondary spectra obtained using smaller separations
were judged to be of dubious validity as they exhibited many emission
features {coincident} with absorption features in the primary
spectrum.

\subsection{The method of Porter et al (2004)}

The model of \citet{JMPorter2004} is based on a constant
convolving function, yet a change in focus along the chip length led
to a change in the width of the flux distribution. To negate this
difficulty the $\mathrm \sigma$ distribution was normalised via a
polynomial fit to remove changes not due to the binary system. The
mean seeing was estimated from the minimum of the now flattened FWHM
spectrum. However, this estimate of the seeing is an upper
estimate. Simulating a binary system with a separation of 0.20\arcsec
and a difference in brightness of 1.8 magnitudes we found the minimum
FWHM may over estimate the seeing by 0.02 pixels. Thus we
subtracted this from our estimate of the seeing using the FWHM
minimum.

\smallskip

\begin{center}
\begin{figure*}
\begin{center}
{\includegraphics[scale=1]{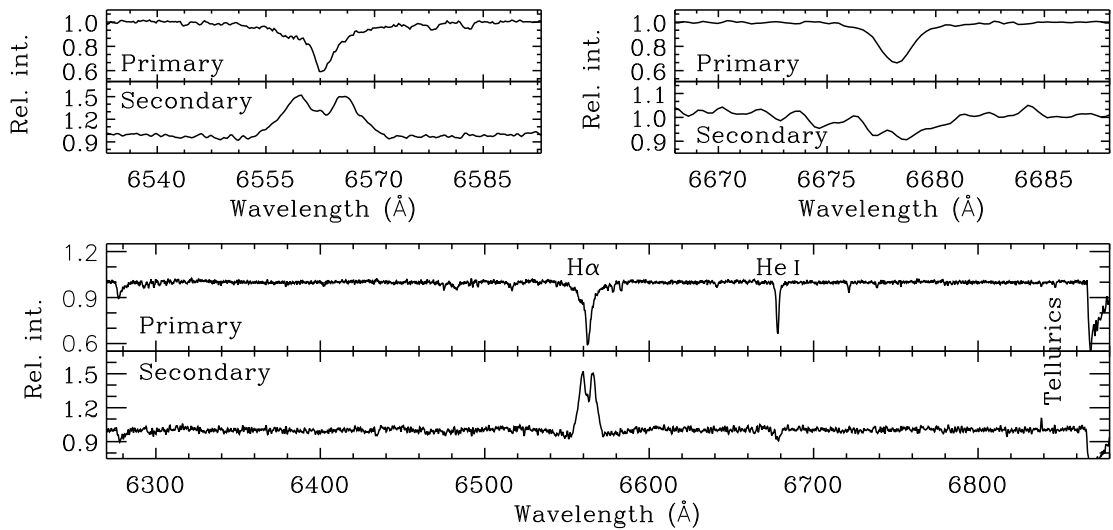}}\\
\end{center}
\begin{center}
\caption{The red spectrum of the close $\mathrm \beta$ Cep binary system, split using the method \citet{JMPorter2004}. In the top left the wavelength range is reduced and centred on $\mathrm H \alpha$ while in the top right attention is focused upon the \ion{He}{i} $\mathrm 6678.15 \AA$ profile. The determined values for the separation and brightness difference were {$d \mathrm = 0.21 \pm 0.07\arcsec$} and {$ \Delta R \mathrm= 1.5 \pm 0.17$}}
\label{porter_spec_split_red}
\end{center}
\end{figure*}
\end{center}

In figure \ref{porter_spec_split_red} we present the red spectrum of
the $\mathrm \beta$ Cep system split using the method of
\citet{JMPorter2004}. The spectrum exhibits a close similarity to the
spectrum obtained using the method of \citet{JBailey1998a} and similar
parameters as those output by the method of \citet{JMPorter2004}.

Applying the method of \citet{JMPorter2004} to the blue spectra did not result in separated spectra, as the model did not converge upon the most likely separation. This could be due to a lack of prominent features in the position spectra. The precision of the blue position spectra was no worse than that in the red spectra (of the order of 1 mas). However, the positional offsets in the blue region are much smaller than in those in the red region as the flux difference in the blue region is less pronounced than that over the H$\rm\alpha$ line.

\smallskip

\section{Results: the separated spectra}

\label{spec_split_res}

\subsection{The spectral types of the binary components}

The spectral type of $\mathrm \beta$ Cep has been estimated to be both
B1 and B2 while the luminosity class of the star has been reported to
be III, IV and V
\citep{Morgan1943,Morgan1955,Lesh1968,Moreletal2006}. These spectral
classifications refer to the total light of the system. As we have
disentangled the spectra of the binary components we can now
spectrally type each component separately. The spectrum of the binary
system has also been designated 'e \& v' where e refers to the
emission nature and v the variability of the spectrum. We assume the
that the primary is responsible for the variability of the spectrum
due to its pulsating nature. {To determine the spectral type of
the binary components we use the spectra that were separated with the
method of \citet{JBailey1998a}, a brightness difference of 1.8
magnitudes and a separation of 0.2$\mathrm''$}.

\smallskip

The presence of \ion{He}{i} absorption in the red spectrum of the
primary indicates a B type star. The equivalent widths of the
H$\mathrm \alpha$ and \ion{He}{i} absorption features suggest that
the primary is an early B (1-3) type star, with a luminosity class of
IV or III. In the blue region the primary spectrum exhibits prominent
\ion{H}{i} absorption lines ($\mathrm H\beta$ and $\mathrm H\gamma$)
alongside \ion{He}{i} absorption lines, which indicates a B type
star. The relatively strong \ion{He}{i} lines and comparatively weak
\ion{H}{i} lines indicate that the primary is an early B type star
(i.e. B0 to B2). Additional features in the primary spectrum include
absorption lines due to \ion{He}{i}, \ion{O}{ii}, \ion{Si}{iii},
\ion{C}{ii}, \ion{C}{iii} and \ion{N}{ii}, lines which are also
suggestive of an early B type star. No \ion{He}{ii} lines are
present, implying the star has a spectral type later than O9. The
ratio of the \ion{He}{i} 4471 $\mathrm \AA$/\ion{Mg}{ii} 4481
$\mathrm \AA$ lines indicate the star is an early B type star, most
probably B2. To determine the luminosity class of the star the ratios
of the \ion{O}{ii} 4415$\mathrm \AA$ \& 4417$\mathrm \AA$ to
\ion{He}{i} 4387$\mathrm \AA$ and \ion{O}{ii} 4348$\mathrm \AA$ to
H$\mathrm \gamma$ lines were used. The strength of the \ion{O}{ii}
lines indicates that the primary's luminosity class is most probably
III.

\smallskip

To summarise, the spectral type of the primary was determined to be
B2III. {The uncertainty associated with this conclusion is
approximately one spectral subtype}.

\smallskip

The most prominent feature in the red spectrum of the secondary is
that of the H$\mathrm \alpha$ emission. The H$\mathrm \alpha$ emission
extends approximately $\mathrm \pm 400$ $ \mathrm {km\,s^{-1}}$ from
the rest wavelength of H$\mathrm \alpha$, as also found by
\citet{RSSchnerr2006}. A Gaussian fit to the $\mathrm H \alpha$
emission profile results in a value for the FWHM of the profile of
$\mathrm \approx$ $\mathrm{500km\,s^{-1}}$. This width is typical of a
rapidly rotating Be star. A \ion{He}{i} $\mathrm 6678 \AA$
absorption feature is present, although weak, with an equivalent width
of $\mathrm \mathrm \approx 0.25 \AA$. This places an
upper limit on the spectral type of about B4/B5.

\smallskip

The secondary spectrum in the blue does not display the many narrow
absorption features due to \ion{O}{ii} and \ion{Si}{iii} the primary
does. Thus there is no reason to suspect it is not a dwarf star. The
spectrum of the secondary in the blue region exhibits absorption lines
due to \ion{H}{i} ($\mathrm H \beta$ and $\mathrm H \gamma$) and
\ion{He}{i} (i.e. $\mathrm 4388, 4471 \AA$). This is indicative of a
B type star. The \ion{He}{i} lines are weaker than than those in the
spectrum of the primary (equivalent widths are approximately 15\%
less) which suggests the secondary is of a later spectral type than
the primary. This is to be expected as it is less bright. In contrast,
the \ion{H}{i} lines in the secondary are of a similar strength to
those in the primary, indicating an early B type star. However, as the
secondary shows emission in H$\mathrm \alpha$ it is possible there is
some emission in H$\mathrm \beta$ and H$\mathrm \gamma$ which 'fills
in' the absorption profile of the secondary. Concentrating on the
strength of the \ion{He}{i} lines, and the ratio of
\ion{He}{i}/\ion{Mg}{ii} (4471/4481$\mathrm \AA$) lines, we
conclude the most likely spectral type of the secondary is B5.

\smallskip

Therefore the spectral type of the secondary is determined to be
B5Ve. {The uncertainty in this spectral typing is again approximately
one spectral subtype}.

\smallskip

{In determining the spectral type of the secondary we assume
that all the flux observed emanates directly from the star in question
and the spectral features observed are purely photospheric in their
origin. To asses the possible flux contribution from circumstellar
material we use the NIR study of \citet{Dougherty1991}. In most cases
mid type Be stars were found to have a small V-J excess, the average
(V-J) excess of a B5Ve star according to \citet{Dougherty1991} is only
$\sim$6\% of the stellar continuum. In addition the continuum excess
falls off very sharply with decreasing wavelengths. Therefore we
expect the continuum excess due to any circumstellar material to be
negligible at optical wavelengths. Regarding emission lines there may
well be unresolved \ion{H}{i} lines in the secondary spectrum
therefore we do not rely on the strength of the \ion{H}{i} lines to
determine the spectral type of the secondary. There may also be some
emission component in the \ion{He}{i} lines. However, in a study of
more than 100 Be stars over a period of 10 years \citet{Chauville2001}
did not observe any Be star with net \ion{He}{i} 4471$\rm\AA$
emission. Therefore, in light of the above comments, we assume that
the \ion{He}{i} lines observed are photospheric in origin.}

\subsection{The mass ratio of the close binary system}

{ Taking the spectral type of the primary to be B2III and that of the
secondary to be B5V and using the tabulated values of \citet{Landolt}
and \citet{Harmanec1988} we obtain values for the masses of
the components of the system of $M_{\mathrm{pri}} \mathrm = 12.6 \pm
3.2 M_{\sun}$ and $M_{\mathrm{sec}} \mathrm = 4.4 \pm 0.7M_{\sun}$
respectively. Uncertainties were estimated by allowing an uncertainty
in the spectral types of one subtype. The mass ratio of the system is
determined to be $\mathrm 0.35 \pm 0.15$.}

\subsection{The $v \sin i$ of the individual binary components}

Using the separated spectra the $v \sin i$ values of each component of
the binary system can be estimated. A rotational profile was
constructed for a variety of $v \sin i$ values{:} from 1 to 550
$\mathrm {km\,s^{-1}}$ in steps of 1 $\mathrm {km\,s^{-1}}$. The
rotation profile was then convolved with synthetic spectra constructed
using ATLAS9 and SYNTHE \citep{Kurucz1993}\footnote{We used the
{GNU Linux} port of ATLAS9 and SYNTHE developed by
\citet{Sbordoneetal2004}. For the initial atmospheric models we used
the grid of models by \citet{Castelli2003}.}. Following the
convolution of the rotational profile and the synthetic spectra the
resultant spectra were broadened by convolution with a Gaussian
function to match the spectral resolution. The rotationally broadened
profile of the \ion{He}{i} 4471 $\mathrm \AA$ line was then compared
with the observed profile using a simple $\mathrm \chi^2$ test.

\smallskip

For the primary synthetic spectra were constructed using values of $
T_{\mathrm{eff}}$ of 23000K, 24000K and 25000K and $\log (g)$ of 3.8.
These values were based on the values determined by
\citet{Catanzaro2008} and the spectral type determined previously. To
simulate the spectrum of the secondary synthetic spectra were
constructed using values of $T_{\mathrm{eff}}$ of 14000K, 15000K,
16000K and 17000K and $\log (g)$ of 4.1. These values are based on a
B5 type dwarf star on the Main Sequence
{\citep{Harmanec1988}}. The range of temperatures used reflects
the uncertainty in the spectral typing. A micro-turbulence of $\mathrm
2$ $\mathrm {km\,s^{-1}}$ was used in the synthetic spectra
generation, and when constructing the rotational profiles a
limb-darkening coefficient of 0.6 was assumed.

\begin{center}
\begin{figure}
\begin{center}
{\includegraphics[scale=0.3]{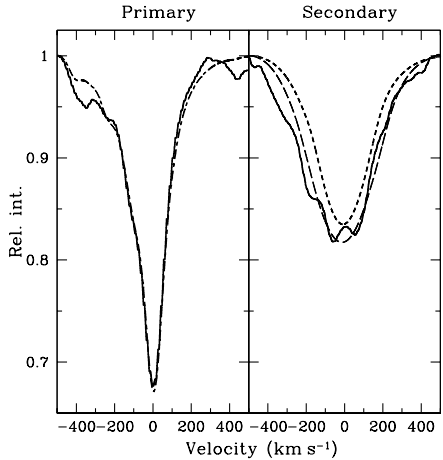}}\\
\end{center}
\begin{center}
\caption{A comparison between the {spectroastrometrically} separated \ion{He}{i} 4471$\mathrm \AA$ profiles and the best fit rotationally broadened synthetic profiles. The solid line represents the \ion{He}{i} line of the primary and secondary spectra respectively, the {short-long} dashed line the best fit primary synthetic spectrum ($T_{\mathrm{eff}}$ 24000K, $\log (g)$ 3.8, $v \sin i$ 29 $\mathrm {km\,s^{-1}}$), the short-dashed line the best fit secondary synthetic spectrum at a $T_{\mathrm{eff}}$ of 15000 K ($\log (g)$ 4.1, $v \sin i$ $\mathrm 163$ $\mathrm {km\,s^{-1}}$) and the long dashed line is the best fit secondary synthetic spectrum at a $T_{\mathrm{eff}}$ of 17000 K ($\log (g)$ 4.1, $v \sin i$ $\mathrm 230$ $\mathrm {km\,s^{-1}}$).}\label{vsinipic}
\end{center}
\end{figure}
\end{center}

The best fit $v \sin i$ values obtained are dependent upon the value
of $T_{\mathrm{eff}}$ used to generate the synthetic spectrum. This is
because the depth of the \ion{He}{I} line increases towards earlier
types. With the previous spectral typing an error in $T_{\mathrm{eff}}$
of $\mathrm \approx 1000$K possible. However, the input $
T_{\mathrm{eff}}$ does not effect the width of the convolved profile. Thus the
most likely value of $ T_{\mathrm{eff}}$ is the one for which the
rotationally broadened synthetic spectrum matches not only the height
but also the width of the observed profile at a given value of $v \sin
i$ (i.e. that associated with the smallest $\mathrm \chi^2$). This
effect is illustrated in Fig.\ \ref{vsinipic} by plotting the best fit
rotationally broadened profiles at 15000K and 17000K over the
secondary profile. The best fit at 15000K fails to fit the
width and depth of the observed profile, which was reflected in the
large value for the minimum $\mathrm \chi^2$ at this
temperature. {The best fit $v \sin i$ values were lower for
lower values of $T_{\mathrm{eff}}$ and ranged from $v \sin i
\mathrm = 132$ $\mathrm km\,s^{-1}$ at $T_{\mathrm{eff}}$ of 14000K to $v \sin i
\mathrm = 230$ $\mathrm km\,s^{-1}$ at $T_{\mathrm{eff}}$ of 17000K. The lowest
$\mathrm \chi^2$ values were obtained with a $T_{\mathrm{eff}}$ of 24000K in
the case of the primary and 17000K in the case of the secondary. We
determine that the primary has a $v \sin i$ of $\mathrm
29^{+43}_{-29}$ $\mathrm {km\,s^{-1}}$ and the secondary has a $v \sin
i$ of $\mathrm 230 \pm 45$ $\mathrm {km\,s^{-1}}$. The
uncertainties in these values were determined from the change in $v
\sin i$ of the synthetic spectra which led to an increase in the
$\chi^2$ of the fit of 1. We note that the secondary profile was best
fit using a B4 star profile ($T_{\mathrm{eff}}$ = 17000K), rather than a B5
profile. This is consistent with an uncertainty of 1 subtype
in the spectral type of the secondary.}

\section{Discussion}
\label{disscussion}

\subsection{The $v \sin i$ of the individual binary components}

The $v \sin i$ of the primary is consistent with literature values of
around 25 $\mathrm {km\,s^{-1}}$ \citep{Teltingetal1997}, although it is
imprecise. This imprecision is due to the use of a relatively wide
slit resulting in an instrumental profile with a FWHM of $\mathrm
\approx 70$ $\mathrm {km\,s^{-1}}$. The $v \sin i$ value for the
secondary is similarly imprecise. However, it is clear that the
secondary does rotate substantially faster than the primary, in
agreement with the estimate of $\mathrm \approx 230$ $\mathrm
{km\,s^{-1}}$ of \citet{Catanzaro2008}.

\smallskip

Given the spectral type of the secondary it's critical velocity is
approximately $\mathrm \approx 430$ $\mathrm {km\,s^{-1}}$
\citep{Townsendetal2004}. It is suggested that the orbit of the
$\mathrm \beta$ Cep system is seen almost edge on \citep[][see also
Sect. 6.2]{Pigulski1992}. If the stars rotate in the plane of the
orbit, {which is not necessarily the case}, then the above $v \sin i$ values are the
intrinsic rotation velocities of the stars. In this case the
rotational velocity of the secondary is 45-65\% of its critical
velocity. This rotation may not be consistent with the hypothesis that
the secondary is a classical Be star.

\smallskip

Whether or not all Be stars rotate at their break-up velocity
is currently a matter or some debate, with arguments both for
\citep{Townsendetal2004} and against \citep{Cranmer2005} this
scenario. However, it is generally accepted that all Be stars rotate
at a substantial fraction of their break-up velocity. To prove
the secondary is rotating at/near its critical velocity we need an
independent determination of its inclination, which is far from
trivial. In addition even if $i$ was constrained, this $v \sin i$
measurement would not be conclusive. It has been shown the method used
here to determine $v \sin i$ returns a lower limit as it does not
account for gravity darkening \citep{Townsendetal2004}. However, the
secondary is shown to be {rotating relatively fast, even at the
lower limit of $i$ = 90$^\circ$, and it may be rotating at a
substantial fraction of its break-up velocity}. This is
{essentially} consistent with the finding of \citet{Porter1996}
who demonstrated Be stars intrinsically rotate at approximately 70\%
of their break-up velocity (using a similar technique to estimate $v
\sin i$). Therefore it is indeed likely the secondary is a classical
Be star.

\subsection{The orbit of $\mathrm \beta$ Cephei}
\label{beta_cep_orbit}

 The orbit of the close companion to $\mathrm \beta$ Cep has been
 determined by \citet{Pigulski1992}. However, both the results
 presented here and by \citet{RSSchnerr2006} suggest a revision of the
 orbital parameters is required. According to the orbit of
 \citet{Pigulski1992} the companion to $\mathrm \beta$ Cep should have
 been to the SW of the primary in 2006. The data presented here, and
 by \citet{RSSchnerr2006} are not consistent with this prediction. As
 the {spectroastrometric} data proves the H$\mathrm \alpha$
 emission is emanating from the close companion, the data place the
 companion to the NE of the primary in 2006.

\smallskip

\begin{center}
\begin{figure}
\begin{center}
\begin{tabular}{c}
{\includegraphics[scale=0.3]{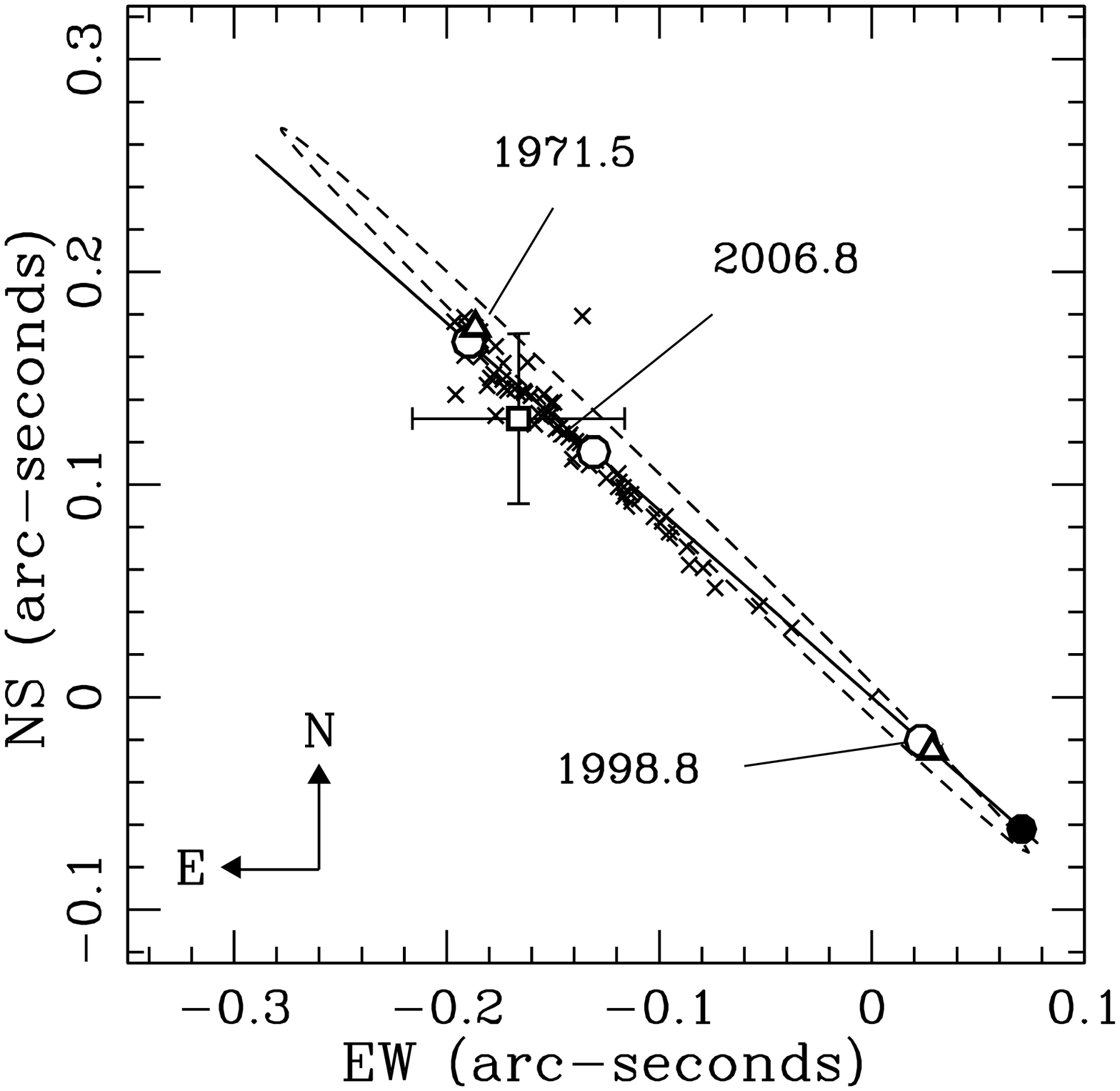}}\\
\end{tabular}
\end{center}
\begin{center}
\caption{\label{new_orbit}A possible relative orbit of the companion
of the close $\mathrm \beta$ Ceph system about the primary. Speckle
interferometric data have been taken from \citet{Pigulski1992} and
\citet{Hartkopfetal2001}, and are marked by crosses. The open square
point represents the position of the companion from our
{spectroastrometric} data, the line indicates the orbital path. We
compare specific orbital predictions and data points at three epochs,
1971.5, 1998.8 and 2006.8. We denote these data points by open
triangles (a square in the case of 2006.8) and orbital predictions by
open circles. The primary is situated at 0,0 and the filled circle
marks the position of periastron. The dashed line shows the orbital
path when $i$ is changed by $\mathrm -2^{\degr}$ and $\mathrm \Omega$
is changed by $\mathrm -3^{\degr}$. The average uncertainty in the
data was approximately 0.015$\mathrm''$ in each direction.}
\end{center}
\end{figure}
\end{center}

Here we investigate whether a slight change of orbital parameters can
result in an orbit consistent with observations. It was found the data
allow a straight line fit, which implies that the system is viewed at
a very high inclination, i.e. $\mathrm \approx 90^{\degr}$. From a
least-squared fit to the data we obtain a value for the position angle
of the line of nodes to be $\mathrm 228.6 \pm 1.4^{\degr}$. We took
the date of 1914.6 as our reference periastron, as did
\citet{Pigulski1992}. For the period of the orbit we considered the
suggestion of \citet{HadravaandHarmanec1996} that the system was
approaching periastron in 1996. This period of 81.4 years differs from
the value used by \citet{Pigulski1992} but is within $\mathrm 3\sigma$
of their value. The semi-major axis of the orbit was estimated via
Kepler's third law, the above period and the total mass of the
system. With the data at our disposal constraining e and $\mathrm
\omega$ is difficult. We set e at 0.6 and investigated what value of
$\mathrm \omega$ was required to fit the observational data, given the
parameters determined above. We found a value of $\mathrm \omega$ of
$\mathrm 20.0^{\degr}$ fit the data well (see Fig.\ \ref{new_orbit}).
\smallskip

The final orbital parameters used were: $\mathrm P_{orb}=81.4yr$,
$\mathrm e=0.60$, $\mathrm T_0=1914.6$, $\mathrm
\Omega=228.6^{\degr}$, $\mathrm i=90.0^{\degr}$, $\mathrm a=0.25''$
and $\omega=20^{\degr}$. Most parameters are within 3$\mathrm \sigma$
of the values of \citep{Pigulski1992}. The orbit is consistent with
previous speckle interferometric observations of the system and our
{spectroastrometrically} determined position of the
secondary. Admittedly, systems are rarely observed at an inclination
of exactly $\mathrm90^{\degr}$. Changing the inclination by up to a
few degrees does little to qualitatively change the picture. Provided
$\mathrm \Omega$ is revised in light of any inclination changes, a
consistent fit to the data is achieved when $i$ is changed by a few
degrees (see Fig \ref{new_orbit}). We stress such an orbit is only
illustrative, and more observations are needed to fully constrain all
the orbital parameters. However, we successfully demonstrate a slight
revision of the orbital parameters of \citet{Pigulski1992} is all that
is required to obtain an orbit consistent with observations.

\section{Conclusions}
\label{conclusion}

In this paper we have studied the close binary system of $\mathrm
\beta$ Cep utilising the novel approach of splitting the binary
spectra using {spectroastrometry}. Disentangling the composite
binary spectrum allows us to determine key properties of each
component. We find that the $\mathrm H \alpha$ emission of the system
is due to the close secondary, as shown by
\citet{RSSchnerr2006}. Splitting the convolved spectrum and assessing
the spectral type of each component we find that the companion is a
dwarf star with a spectral type of B5. We also find {it may
rotate at a substantial fraction of its critical velocity, with a lower
limit of $v \sin i$ = 230 $\mathrm {km\,s^{-1}}$ corresponding to
$V_{\mathrm{r}}/V_{\mathrm{crit}}$ of 53\%}.

\smallskip

The secondary's estimated mass and $v \sin i$ fall within the range of
typical values for classical Be stars. Thus we consider it highly
likely the secondary is indeed such a star. Therefore we have not only
confirmed the result of \citet{RSSchnerr2006} but we validate their
suggestion that the secondary could be a classical Be star. In which
case the H$\mathrm \alpha$ emission is thought to be due to gaseous
equatorial material ejected from the star by a combination of rapid
rotation and some other phenomena, e.g. non-radial pulsation
\citep{PorterandRivinius2003}. As the H$\mathrm \alpha$ emission is
shown to originate from a star where the standard Be paradigm applies
the $\mathrm \beta$ Cep system does not pose any contradictions to the
current understanding of either $\mathrm \beta$ Cep stars or classical
Be stars.

\begin{acknowledgements}

 The William Herschel Telescope is operated on the island of La Palma
by the Isaac Newton Group in the Spanish Observatorio del Roque de los
Muchachos of the Instituto de Astrofísica de Canarias. R.D.O. is
grateful for the support from the Leverhulme Trust for awarding a
Research Fellowship.  H.E.W gratefully acknowledges a PhD studentship
from the Science and Technology Facilities Council of the United
Kingdom (STFC). {The authors wish to thank an anonymous referee
for a careful reading of the manuscript and insightful comments which
helped improve the paper.}

\end{acknowledgements}

\bibliographystyle{aa}
\bibliography{bib.bib}
\end{document}